\documentclass[twocolumn,aps,pra,superscriptaddress,nofootinbib]{revtex4-2}

\usepackage{amsmath,amsthm}
\usepackage{graphicx}
\usepackage{hyperref}

\hypersetup{
  colorlinks   = true, 
  urlcolor     = blue, 
  linkcolor    = blue, 
  citecolor    = red   
}

\newcommand{\N}{{\mathcal{N}}}

\newtheorem{definition}{Definition}

\begin{document}

\title{Rare Events Are Nonperturbative:\\ Primordial Black Holes From Heavy-Tailed Distributions}

\author{Sina Hooshangi}
\email{sina.hooshangi@ipm.ir}
\affiliation{School of Astronomy, Institute for Research in Fundamental Sciences (IPM), Tehran, Iran, P.O. Box 19395-5531}

\author{Mohammad Hossein Namjoo}
\email{mh.namjoo@ipm.ir}
\affiliation{School of Astronomy, Institute for Research in Fundamental Sciences (IPM), Tehran, Iran, P.O. Box 19395-5531}

\author{Mahdiyar Noorbala}
\email{mnoorbala@ut.ac.ir}
\affiliation{Department of Physics, University of Tehran, Iran. P.O.\ Box 14395-547}
\affiliation{School of Astronomy, Institute for Research in Fundamental Sciences (IPM), Tehran, Iran, P.O. Box 19395-5531}
 

\begin{abstract}
In recent years it has been noted that the perturbative treatment of the statistics of fluctuations may fail to make correct predictions for the abundance of primordial black holes (PBHs). Moreover, it has been shown in some explicit single-field examples that the nonperturbative effects may lead to an exponential tail for the probability distribution function (PDF) of fluctuations responsible for PBH formation---in contrast to the PDF being Gaussian, as suggested by perturbation theory. In this paper, we advocate that the so-called $\delta N$ formalism can be considered as a simple, yet effective, tool for the nonperturbative estimate of the tail of the PDF. We discuss the criteria a model needs to satisfy so that the results of the classical $\delta N$ formalism can be trusted and most possible complications due to the quantum nature of fluctuations can be avoided. As a proof of concept, we then apply this method to a simple example and show that the tail of the PDF can be even {\it heavier} than exponential, leading to a significant enhancement of the PBH formation probability, compared with the predictions of the perturbation theory. Our results, along with other related findings, motivate the invention of new, nonperturbative methods for the problem and open up new ideas on generating PBHs with notable abundance.
\end{abstract}

\maketitle

\section{Introduction}

The observation of gravitational waves from binary systems by LIGO/Virgo \cite{LIGOScientific:2016aoc}, LIGO/VIRGO/KAGRA \cite{LIGOScientific:2021djp}  and the possibility that the 
PBHs 
 are
the source of these events has revived interest in the study of PBHs \cite{Sasaki:2016jop, Clesse:2016vqa, Bird:2016dcv, Carr:2016drx, Sasaki:2018dmp, Biagetti:2018pjj, Martin:2019nuw, Fumagalli:2020adf, Carr:2020xqk, Carr:2020gox}.  The origin of the PBHs is large curvature perturbations $\zeta\gtrsim1$ that enter the horizon during the radiation era in the early universe and collapse an entire horizon-sized region into a black hole \cite{Hawking:1971ei,Carr:1974nx, Carr:1975qj, Chapline:1975ojl}.  Such fluctuations are known to be generated during inflation with a 
PDF $\rho_\zeta$, to be determined by the physics governing inflation.  The abundance of produced PBHs is then characterized by
\begin{equation}\label{beta-zeta}
\beta = \int_{\zeta_c}^\infty \rho_\zeta d\zeta,
\end{equation}
where $\zeta_c$ is some critical threshold that has to be calculated from collapse models \cite{Carr:2020gox}.  Here we ignore the subtleties associated with the determination of $\zeta_c$ and simply assume $\zeta_c=1$.

A major obstacle in computing $\beta$ is the difficulty in calculating $\rho_\zeta$ for large values of $\zeta$, i.e., at its tail.  At a na\"ive level, one may assume $\rho_\zeta$ is Gaussian (as predicted for small perturbations and as observed on the CMB) and extrapolate the observed power spectrum of curvature perturbations on CMB scales to PBH scales.  This leads to minuscule PBH abundance for most inflationary models, unless exotic features are added to the potential to enhance the power at shorter scales.  However, the Gaussian approximation is clearly unjustified away from the peak of the PDF and a reliable calculation of $\rho_\zeta$ at its tail is necessarily of nonperturbative nature.  There have been a number of nonperturbative investigations taking into account various effects that give rise to an enhancement in the PBH abundance, sometimes by several orders of magnitude \cite{Pattison:2017mbe, Panagopoulos:2019ail, Ezquiaga:2019ftu, Ballesteros:2020sre, Celoria:2021vjw, Biagetti:2021eep, Pattison:2021oen, Figueroa:2020jkf, Figueroa:2021zah, Cohen:2021jbo}.  This enhancement is due to lifting the Gaussian fall-off of $\rho_\zeta$ to a slower type of decay; the slowest single-field case being an exponential fall-off $e^{-\Lambda\zeta}$.  This raises a natural question whether a heavier tail for $\rho_\zeta$, slower than any exponential, is possible.  We show that a heavy tail is indeed possible.  To achieve this, we employ the $\delta N$ formalism \cite{Sasaki:1995aw, Wands:2000dp, Lyth:2004gb, Sugiyama:2012tj, Abolhasani:2019cqw}, which is much simpler than other approaches used elsewhere.  Unlike the conventional usage of the $\delta N$ formalism in other applications, which is based on Taylor expanding $N$ in small perturbations, we work with the full nonlinear $N$ and study how this affects the PDF of $\zeta$ and modifies its tail. This nonperturbative view of the $\delta N$ formalism is permitted 
since it
holds nonlinearly for large-scale fluctuations. We shall argue that there will be conditions under which our approach is valid and other effects---not captured by the classical $\delta N$ formalism---can be neglected.  We will discuss these conditions in general and for a toy model example that we use to demonstrate our idea. Throughout this paper,  we set $M_P = 1/\sqrt{8\pi G_N} = 1$.


\section{Outline of the Method}
 \label{sec:outline}
 
According to the $\delta N$ formalism, the curvature perturbation $\zeta$ on a final uniform density surface is related to the perturbation $\delta\phi$ on an initial surface by
\begin{equation}\label{deltaN}
\zeta = \N(\bar\phi + \delta\phi) - \N(\bar\phi),
\end{equation}
where $\N(\phi)$ is the  total number of $e$-folds (in an unperturbed universe) between the initial hypersurface, labeled by the field value $\phi$, and the final uniform density hypersurface; $\bar\phi$ is the unperturbed initial value of the field, and the mode under consideration is super-horizon 
 throughout
this evolution.  The conventional approach is to treat this equation perturbatively and make Taylor expansions to first order 
\begin{equation}\label{deltaN-O1}
\zeta \approx \N'(\bar\phi) \delta\phi,
\end{equation}
or 
second order
\begin{equation}\label{deltaN-O2}
\zeta \approx \N'(\bar\phi) \delta\phi + \frac12 \N''(\bar\phi) \delta\phi^2.
\end{equation}
The PDF of the field perturbations $\delta\phi$ on the initial surface is often assumed to be Gaussian:
\begin{equation}\label{PDF-delta-phi}
\rho_{\delta\phi} = \frac{1}{\sqrt{2\pi} \sigma_{\delta\phi}} e^{-\delta\phi^2/2\sigma_{\delta\phi}^2},
\end{equation}
where $\sigma_{\delta\phi}$ is equal to $H/2\pi$ at the initial time.  It then follows that, to first order, the PDF of $\zeta$ is also Gaussian with $\sigma_{\zeta} = |\N'(\bar\phi)|\sigma_{\delta\phi}$.  Including the second order corrections, one finds a non-Gaussianity $\propto \N''/\N'^2$.  We don't appeal to such Taylor expansions in this paper; instead use the fully nonlinear relation between $\zeta$ and $\delta\phi$.  In particular, the PDF of $\zeta$ reads
\begin{equation}\label{PDF-zeta-vs-delta-phi}
\rho_\zeta = \left| \frac{d\delta\phi}{d\zeta} \right| \rho_{\delta\phi} = \frac{1}{|\N'(\bar\phi + \delta\phi)|} \rho_{\delta\phi},
\end{equation}
where, for simplicity, we have assumed that $\N(\bar\phi)$ is a one-to-one function. 
An important consequence of this nonlinearity is that large values of $\zeta$ need not be mapped to large values of $\delta\phi$.  Therefore, the tail of $\rho_\zeta$ does not need to inherit the Gaussian nature of the tail of $\rho_{\delta\phi}$.  This is a key point that plays a central role in our approach.

 Given a potential $V(\phi)$, our strategy to calculate the PDF of $\zeta$ at a given $\bar\phi$, corresponding to a given PBH scale, is as follows:
We assume that $\rho_{\delta\phi}$ is Gaussian on the initial surface at $\bar\phi$.  Next we calculate $\N$ without resort to any slow-roll approximation.\footnote{ Since the nontrivial evolution of $\zeta$ in single-field models takes place in a non-attractor phase where the slow-roll conditions are violated, it is essential to avoid slow-roll approximations.}  
To do so, we use the equation of motion
\begin{equation}\label{KG}
\phi'' + \left( 3  - \frac12 \phi'^2 \right) \left( \phi' + \frac{V_{,\phi}}{V} \right) = 0,
\end{equation}
where prime denotes derivative with respect to the number of $e$-folds defined by $dN=H dt$.\footnote{Throughout this
paper, a prime on a function denotes differentiation with respect to
its natural argument. So ${\cal N}'$ means $d{\cal N}/d\phi$ whereas
$\phi'$ means $d\phi/dN$.}
 We define the final hypersurface (after which $\zeta$ remains intact until the horizon reentry during the radiation era) by $\phi=\phi_e$.
We can then find $\N(\bar \phi+\delta \phi)$ by solving 
 Eq.~\eqref{KG}
with initial conditions $\bar \phi+\delta \phi$ and $\bar\pi = \bar \phi'$ for a range of different values of $\delta \phi$ and then requiring that $\phi(N = \N(\bar \phi+\delta \phi)) = \phi_e$.  This yields $\N$ as a function of the initial values $\bar \phi+\delta \phi$, which can then be inserted in Eq.~\eqref{PDF-zeta-vs-delta-phi}, in conjunction with \eqref{PDF-delta-phi}, to provide the PDF of $\zeta$. Notice that the initial field velocity will not be perturbed (while, in principle, $\N$ does depend on both initial conditions). This will be justified 
 later
after 
implementing
the method in a simple model.

In the next section we work out the details for a particular potential 
 leading
to a heavy-tailed PDF. Here, we clarify what we mean by a heavy tail (which is consistent with the standard nomenclature in mathematics)~\cite{FKZ}:
\begin{definition}
A PDF $\rho_\zeta(\zeta)$ is said to be heavy-tailed, if its tail function $\bar F(\zeta) = \int_\zeta^{+\infty} \rho_\zeta(\zeta') d\zeta'$ beats any exponentially decaying function, i.e., if for every $\lambda>0$, we have $\limsup_{\zeta\to+\infty} e^{\lambda\zeta} \bar F(\zeta) = +\infty$.
\end{definition}
This is not very easy to check in a numerical calculation, so we employ a related definition that is more practical:
\begin{definition}\label{practical-def}
We call a PDF $\rho_\zeta(\zeta)$ practically heavy-tailed if $\lim_{\zeta\to+\infty} {\cal D} = 0$, where ${\cal D} = -d\log\rho_{\zeta}/d\zeta$.
\end{definition}
Note that both definitions exclude exponential tails (where $\cal D$ approaches a positive constant) and of course Gaussian tails, but admit power-law tails, as well as those falling like $\exp[-k\zeta^p]$ with $0<p<1$ and $k>0$.

\section{A Toy Model}\label{sec:toy}

As a proof of concept we carry out the above procedure for a simple toy model.
Let us suggest the potential
\begin{equation}\label{V}
V(\phi) = V_0 \left[ 1 + \frac13 \alpha \phi^3 \right],
\end{equation}
and choose the unperturbed initial value to be $\bar\phi=0$ (equivalently, one may replace $\phi$ by $\phi-\bar\phi$ throughout).  We could work with a number of other inflationary potentials, but we can offer  a heuristic derivation for this particular
$V$ based on requiring a power-law tail for the PDF of $\zeta$, that turns out to be of the form $\rho_\zeta \sim \zeta^{-2}$ (hence ${\cal D} = 2/\zeta$), starting from a Gaussian PDF for $\delta \phi$.
See 
Appendix~\ref{appendix} for the details of this derivation.

\begin{figure}
\includegraphics[scale=.7]{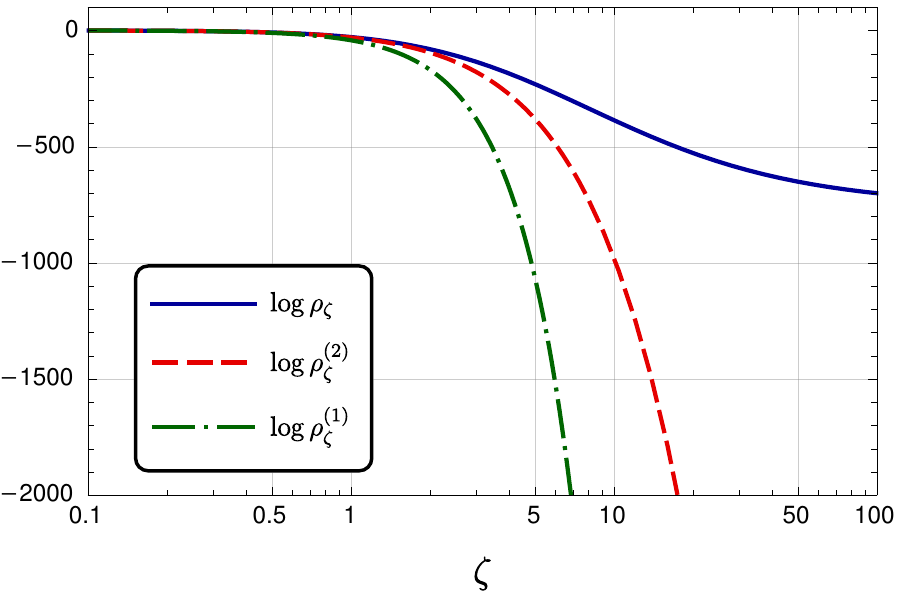}
\caption{The PDF $\rho_\zeta$ for our toy model computed by the fully nonlinear $\delta N$ formula (solid blue), the linear approximation (dashed-dotted green), and the quadratic approximation (dashed red).  We have $\bar\phi = 0$, $\bar\pi = -10^{-4}$, $\phi_e = -3.5\times10^{-4}$, $\alpha = 10^4$, $V_0 = 10^{-10}$ corresponding to $\sigma_{\delta\phi} = 9.2\times10^{-7}$.}
\label{fig:PDFs}
\end{figure}

We now 
follow the method outlined in the previous section to obtain the PDF associated with the potential~\eqref{V}
by choosing numerical values for $\bar\phi>\phi_e$ and for the range $\delta\phi_{\rm min} < \delta\phi < \delta\phi_{\rm max}$.
Importantly, notice that we assume, for simplicity, an abrupt transition to a slow-roll phase of inflation immediately after $\phi_e$. As argued in \cite{Namjoo:2012aa} and further investigated in \cite{Cai:2018dkf} this sharp transition leads to the conservation of $\zeta$ just after the transition so that following its evolution up to the end of inflation will not be needed. It would be interesting to study the effect of a milder transition to our results but it is beyond the scope of this paper.  We require Eq.~\eqref{V} to be valid only in the range $\phi_e < \phi < \bar\phi + \delta\phi_{\rm max}$; beyond that, it suffices that a slow-roll phase be realized.  Furthermore, notice that we choose an initial 
velocity that leads to a non-attractor phase of inflation from $\phi$ to near $\phi_e$. This 
phase leads to the nontrivial evolution of $\zeta$ on super-horizon scales, leading to a non-standard PDF for $\zeta$. 

In Fig.~\ref{fig:PDFs} we show $\rho_\zeta$ as a result of the procedure outlined above.  In addition to plotting the fully nonlinear result $\rho_\zeta$, we have computed $\N'(\bar\phi)$ and $\N''(\bar\phi)$ numerically, inserted them in Eqs.~\eqref{deltaN-O1} and \eqref{deltaN-O2}, and followed the rest of the numerical calculation again using Eq.~\eqref{PDF-zeta-vs-delta-phi} but now with these linear and quadratic relations between $\zeta$ and $\delta\phi$.  This gives $\rho_\zeta^{(1)}$ and $\rho_\zeta^{(2)}$, respectively, which depart from $\rho_\zeta$ around $\zeta\sim1$ and are markedly suppressed afterwards.  The values of the parameters used in this calculation are quoted in the caption of Fig.~\ref{fig:PDFs}.

\begin{figure}
\includegraphics[scale=.25]{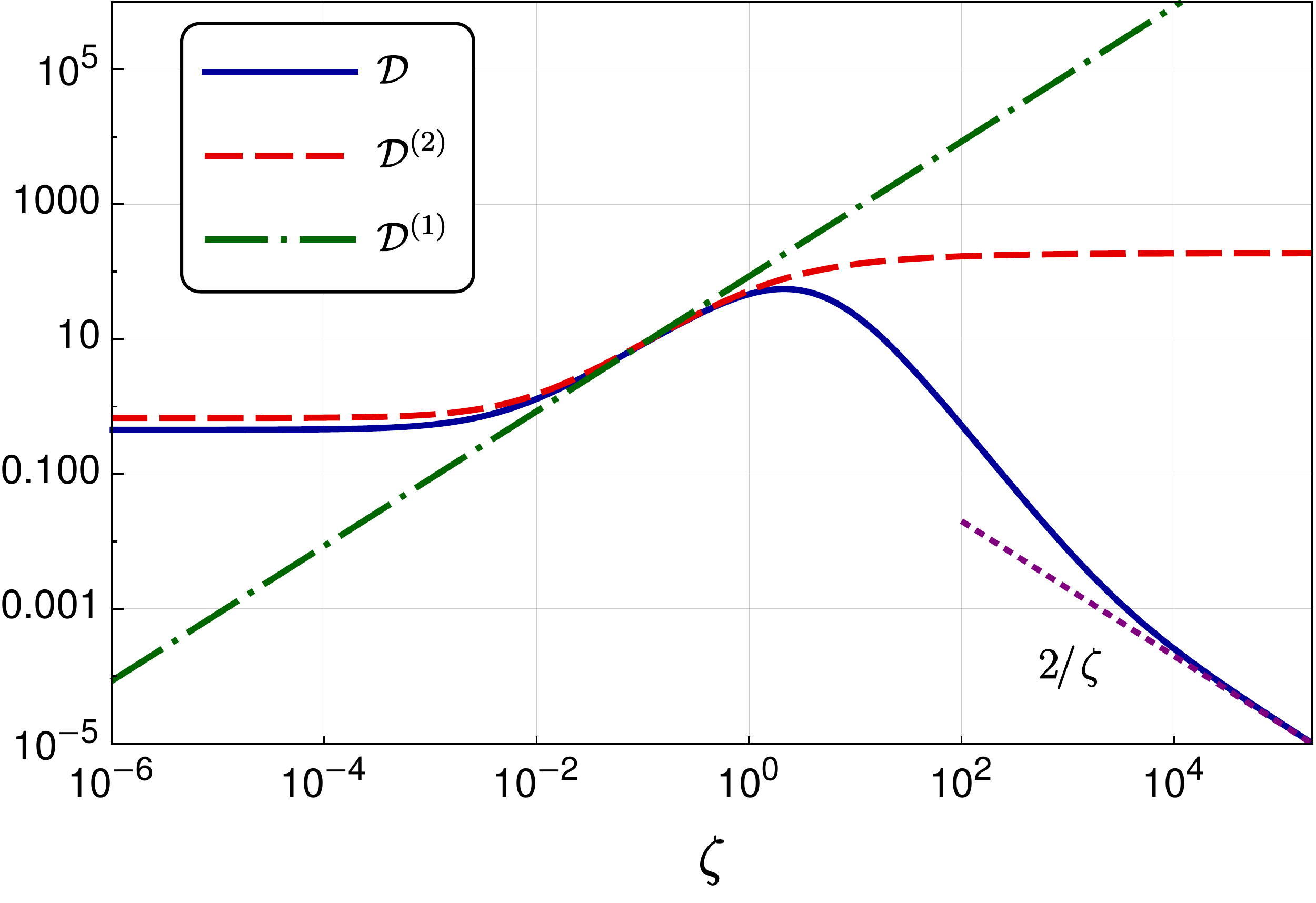}
\caption{The practical measure of heaviness of the PDF, ${\cal D}$ defined in Def.~\ref{practical-def}, for our toy model; same legend and parameters as Fig.~\ref{fig:PDFs}.  The dotted oblique line in the bottom right corner is the analytical prediction 
for large $\zeta$.}
\label{fig:der-PDFs}
\end{figure}

Fig.~\ref{fig:der-PDFs} is intended to show the heaviness of the tail by plotting 
$\cal D$, defined in Def.~\ref{practical-def}.  Again, we have demonstrated the result for ${\cal D}^{(1)}$ and ${\cal D}^{(2)}$, too.  The nonstop fall-off trend in ${\cal D}$ is a plausible witness for the heaviness of its tail.  We are further assured by the perfect agreement between the predicted analytic behavior ${\cal D} = 2/\zeta$ and the numerical result.  On the other hand, the linear growth of ${\cal D}^{(1)}$ indicates the Gaussian decay of $\rho^{(1)}$.  Similarly, ${\cal D}^{(2)}$ approaches a nonzero constant, 
so $\rho^{(2)}$ has an exponential tail, which is not heavy either.  The fact that $\rho^{(2)}$ decays exponentially is a 
consequence of the quadratic nature of Eq.~\eqref{deltaN-O2}:  For large $\zeta$ we have $\zeta \propto \delta\phi^2$ and so the exponent in Eq.~\eqref{PDF-delta-phi} becomes linear in $\zeta$.

\begin{figure}
\includegraphics[scale=.24]{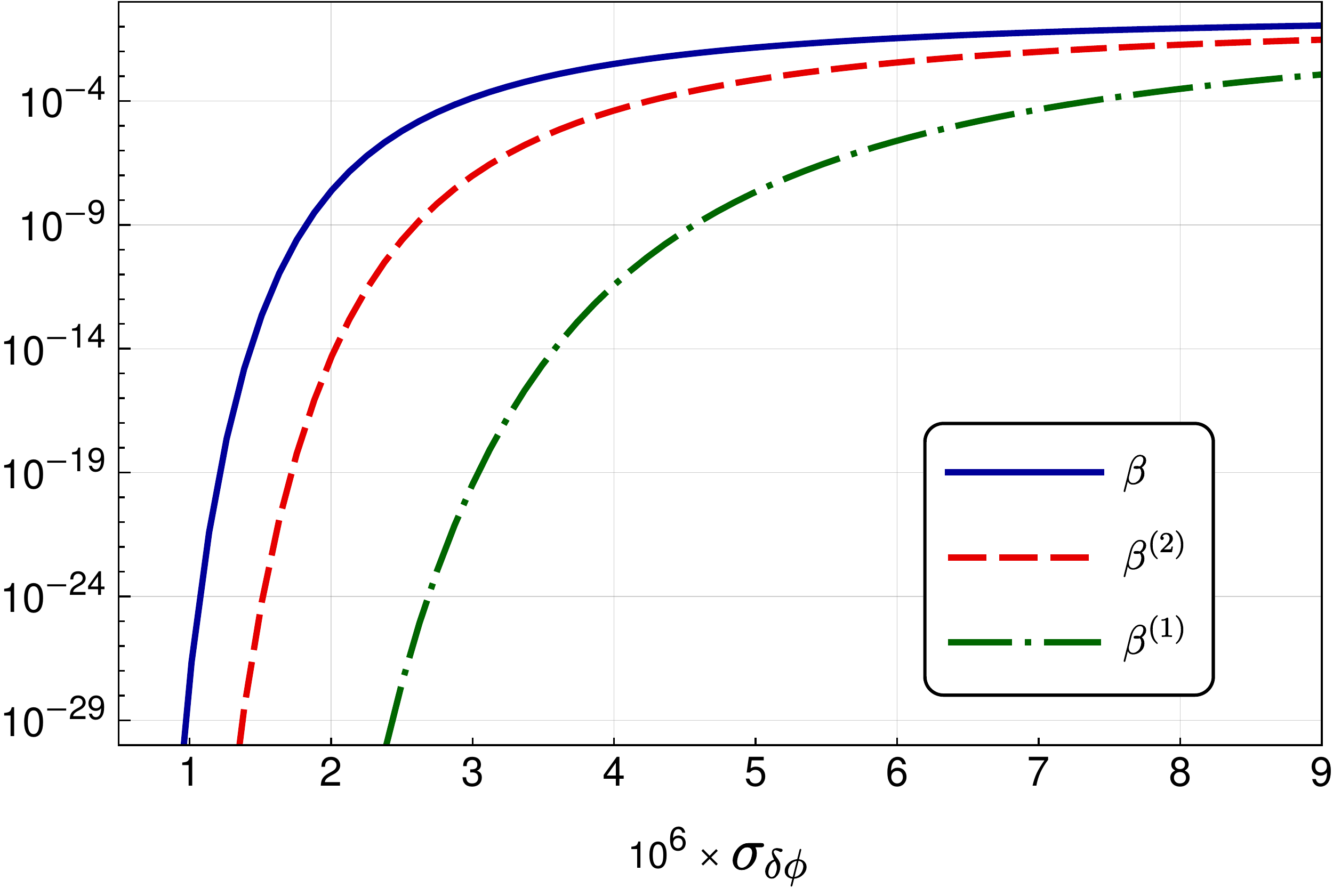}
\caption{The PBH abundance parameter $\beta$ for our toy model as a function of $\sigma_{\delta\phi}$; same legend and parameters as Fig.~\ref{fig:PDFs}, except $V_0$ which varies with $H$ and we change $\phi_e$ to $-2.5 \times 10^{-5}$ to obtain larger separation between curves.}
\label{fig:beta}
\end{figure}

The parameter $\beta$ in Eq.~\eqref{beta-zeta}, which is related to the PBH abundance, is plotted in Fig.~\ref{fig:beta} as a function of $\sigma_{\delta\phi} = H/2\pi$.  We have also plotted $\beta^{(1)}$ and $\beta^{(2)}$, which are derived from $\rho^{(1)}$ and $\rho^{(2)}$, respectively.  We see that taking into account the full nonlinear nature of the $\delta N$ formula \eqref{deltaN} amounts to considerable enhancement in the PBH abundance compared to the perturbative calculation $\beta^{(1)}$ based on the Gaussian PDF $\rho_\zeta^{(1)}$.  One may argue that maybe it was unnecessary to incorporate the full nonlinear form of Eq.~\eqref{deltaN} and that if one included the second order term as in Eq.~\eqref{deltaN-O2}, the same level of enhancement could be reached by $\beta^{(2)}$.  However this is not the case in the 
 present model,
as Fig.~\ref{fig:beta} clearly shows that $\beta^{(2)} \ll \beta$ by several orders of magnitude.  We conclude that the result, including the enhancement, is intrinsically nonperturbative. We further note that the large $\zeta$ behavior of 
$\rho_\zeta$
while consistent with analytical predictions, does not suffice to estimate $\beta$ since the integral in Eq.~\eqref{beta-zeta} is mainly supported 
around
$\zeta \sim {\cal O}(1)$. As a result---just like the $\zeta \ll 1$ limit---the methods that only predict the behavior of the PDF in the $\zeta \gg 1$ limit may not reliably estimate $\beta$. It is  indeed the transient regime between the two limits that contributes to $\beta$, which might be considered as a serious technical difficulty. This is not an issue here as we were able to construct the full PDF using the $\delta N$ formalism. However, note that 
our approach
relies on some hidden assumptions that 
need to be
satisfied. We will briefly discuss these assumptions in the next section and confirm that they are indeed met in our presented example.

\section{Underlying Assumptions}\label{sec:assumptions}

In this section we justify the various underlying assumptions and approximations made in our work.

\textbf{Non-Gaussianities in $\delta \phi$:} We have assumed that $\delta\phi$ has a Gaussian PDF.  Although it is straightforward to incorporate a non-Gaussian PDF into our numerical calculations, the precise form of such a PDF requires an in-in calculation and depends on the exact form of the interactions dictated by higher order terms ${\cal L}_n$ ($n>2$) in the Lagrangian of $\delta\phi$.  If these higher order (interaction) terms are negligible compared to the second order (free) term ${\cal L}_2$, i.e., if ${\cal L}_n \ll {\cal L}_2$ for $n>2$, we can safely ignore the non-Gaussianity of $\rho_{\delta\phi}$ for the range of $\delta\phi$ under consideration.  To this end, we estimate $\dot{\delta\phi}$ and $\frac1a \nabla\delta\phi$ by $H\delta\phi$, so that a typical term in the Lagrangian is of order ${\cal L}_n \sim c_n(t) \delta\phi_{\rm max}^n$ where $\delta\phi_{\rm max}$ corresponds to $\zeta_{\rm max}$, the largest fluctuation that is relevant to our calculation of the PBH abundance.  The ratio of interest is therefore ${\cal L}_n / {\cal L}_2 \sim (c_n/c_2) \delta\phi_{\rm max}^{n-2}$ which we need to ensure remains small. In practice, it often suffices to check the size of ${\cal L}_3/{\cal L}_2$ and in most situations it is the matter sector---rather than the gravity sector---that 
contributes
most to the interactions, reducing this ratio to $|V'''/V| \delta \phi_{\rm max}$.

\textbf{Horizon Crossing Effects:} There is a 
common lore that the horizon crossing effects cause inaccuracies in the $\delta N$ formalism of the order of the slow-roll parameters.  Accordingly, a na\"ive application of the $\delta N$ formalism does not yield the non-Gaussianity consistency relation $f_{\rm NL} \propto (n_s-1)$ of single-field inflation \cite{Maldacena:2002vr}.  It may therefore be a source of concern that such effects may as well spoil our nonperturbative calculation.  However, it is not true that these horizon crossing effects render the $\delta N$ formalism invalid.  They do cause corrections, but those corrections appear in the correlations of $\delta\phi$ at the initial surface, whose deviation from Gaussianity is controlled by interaction terms ${\cal L}_n$ as we mentioned in the previous paragraph.\footnote{In fact, as a consistency check of this statement it can be shown that a careful calculation of these corrections reproduces the well-known consistency relation within the $\delta N$ formalism \cite{Abolhasani:2018gyz}.} Therefore, the smallness of ${\cal L}_{n} / {\cal L}_2$ can simultaneously resolve such concerns.

\textbf{Stochastic Effects:} We have neglected stochastic effects throughout this paper.  Indeed there are two sources of randomness that give rise to a PDF for $\zeta$.  First, there is the randomness of the initial conditions $\delta\phi$ of the field at the initial surface, which is incorporated as $\rho_{\delta\phi}$ in our calculation.  The second source is the stochastic noise that the field encounters along the way.  Let us consider how a rare event $\zeta\gtrsim1$ can be generated under the effect of these sources of randomness.  One possibility is to start with a rare initial condition (from the tail of $\delta\phi$) but then proceed on a typical trajectory (close to the deterministic evolution of the field) toward the end of inflation.  Another possibility is to start with a typical initial field value (around the peak of $\rho_{\delta\phi}$) but then follow a rare path (highly fluctuating and far from the average trajectory) until the end of inflation.  It is highly unlikely that both extreme events occur simultaneously in a single realization, so it is fair to say that the probability of a rare event $\zeta\gtrsim1$ is the sum of the probabilities of these two alternative possibilities.

We have only considered the first possibility in this work, so strictly speaking, we have a lower bound on the PDF of $\zeta$ and hence on the PBH abundance.  This is already a considerably enhanced probability compared to the Gaussian or exponential suppression.  To justify the approximation that the inflaton proceeds on a typical path, let us estimate the size of the stochastic noise when diffusion is small.  It is well known that the stochastic noise corresponds to random jumps of magnitude $H/2\pi$ per $e$-fold that are superimposed on top of the classical equation of motion of $\phi$.  Let $\Delta\phi$ be the total field excursion in a given path from the initial value $\bar\phi + \delta\phi$ to $\phi_e$.  The expected displacement due to the accumulated jumps is then, just as in the conventional random walk, equal to $(H/2\pi)\sqrt\N$, where $\N$ is the total number of $e$-folds during this excursion.  Therefore, a fair measure of relative importance of the stochastic effects is given by
\begin{equation}\label{stochasticity}
{\cal S} = (H/2\pi\Delta\phi)\sqrt\N;
\end{equation}
so we need to check that $\cal S$ is small.

We have not considered the second possibility (following a rare path) here.  But other studies \cite{Pattison:2017mbe, Ezquiaga:2019ftu, Celoria:2021vjw, Tada:2021zzj} claim no more than an exponential enhancement due to the stochastic effects.  It is therefore reasonable to assume that the main contribution to the PBH abundance comes from our approach. 

Finally, note that $\bar\pi$ must be chosen such that the inflaton classically reaches $\phi_e$ for all $\delta\phi \in [\delta\phi_{\rm min}, \delta\phi_{\rm max}]$.


\textbf{Contribution of $\delta\pi$:} 
We have ignored the fluctuations to the conjugate momentum $\pi = \phi'$,
so we work with a fixed value of $\phi'_0$  set equal to $\bar\phi'$.  This is justified whenever the mode function $\delta\phi$ freezes and hence $\delta\pi$ decays exponentially. Since $\delta \phi$ is the perturbation of an 
almost massless field
during inflation, this assumption is well justified.

\begin{table}
\begin{tabular}{|c||c|c|}
\hline 
& $\zeta_{\rm max} = 1$ & $\zeta_{\rm max} = 10^5$ \\
\hline\hline
${\cal L}_3 / {\cal L}_2$ & 0.14 & 0.71 \\
\hline
$\cal S$ & 0.0060 & 0.75 \\
\hline 
\end{tabular}
\caption{The parameters ${\cal L}_3 / {\cal L}_2$ and $\cal S$ for our model with parameters given in Fig.~\ref{fig:PDFs}.}\label{tab:data}
\end{table}

Before closing this section, let us verify the validity of these assumptions in the toy model of the previous section.  Let us first consider the set of parameters used in Figs.~\ref{fig:PDFs} and~\ref{fig:der-PDFs}.  In the single-field model at hand, there are 11 terms ${\cal L}_3$ in the cubic Lagrangian in the flat gauge each with its own $c_3$ \cite{Maldacena:2002vr}. We quote the maximum of the ratios ${\cal L}_3 / {\cal L}_2$ for two values of $\zeta_{\rm max}$ in Table~\ref{tab:data}.  For each individual path with initial condition $\delta\phi$, we calculate $\Delta\phi$ and $\N$. The maximum 
$\cal S$ among all paths with $\zeta<\zeta_{\rm max}$ in our numerical calculation is given in Table~\ref{tab:data}.  It is remarkable that both ${\cal L}_3 / {\cal L}_2$ and $\cal S$ are small, even for $\zeta$ as large as $10^5$.  Thus $\rho_\zeta$ is reliable over the entire visible range in Figs.~\ref{fig:PDFs} and \ref{fig:der-PDFs}---including the heavy tail.  Fig.~\ref{fig:beta} uses a different set of parameters for which our $\rho_\zeta$ at the tail is not as reliable.  However, those values are not relevant for the purpose of calculating $\beta$; instead we can confirm that ${\cal L}_3 / {\cal L}_2 \approx 0.22$, and $\cal S$ varies between $0.02$ to $0.31$ for $\zeta \sim 1$ which is the region relevant to $\beta$.  So we see that in all cases the approximations are justified.

\section{Concluding Remarks}\label{sec:conclusions}

Our primary point has been to give a proof of concept. There are several issues that need to be addressed to calculate the PBH abundance reliably. To begin, the Eq.~\eqref{beta-zeta} for $\beta$ is only a crude estimate. A more accurate treatment requires more rigorous determination of the threshold \cite{Musco:2018rwt, Young:2019yug, Musco:2020jjb}. Actually, the physically relevant quantity is the density contrast $\delta$ rather than $\zeta$. These subtleties modify our estimate of $\beta$, but not the fact that it differs significantly from the predictions of perturbation theory. The second issue is the contributions of the stochastic effects, which as we discussed, can further lift the PDF, albeit not as heavily as we have found. As we mentioned before, we have not considered
these effects by working in the drift-dominated regime, but they
deserve further study. In particular, it would be interesting to
explore scenarios where both classical and stochastic effects
(corresponding, respectively, to the initial and on-the-way
fluctuations) are equally important.

There are various extensions that can be made to this work. In a future publication, we will apply the nonperturbative $\delta N$ to other single-field models to see how generic the heavy tail might be and also to investigate how $\beta$ may be modified compared to the perturbative treatments. As another extension, one can apply the same methods to multi-field models of inflation \cite{Hooshangi:2022lao}.
Furthermore, the possibility of obtaining heavy tails in multi-field scenarios seems to be fairly unexplored (see, however, Ref.~\cite{Panagopoulos:2019ail}).
It is also interesting to investigate if the contribution of the tail region to $\beta$ can be made more pronounced. 

\begin{acknowledgments}
We thank Eva Silverstein for useful comments.  M.N.\ acknowledges financial support from the research council of University of Tehran.
\end{acknowledgments}

\appendix

\section{
	     Speculating a potential leading to a heavy tail}\label{appendix}
	
Here we briefly outline the analysis that suggests the potential \eqref{V}, leading to a power-law tail for $\rho_\zeta$.
For large values of $\zeta$, we like to avoid the Gaussian fall-off due to $\rho_{\delta\phi}$ in Eq.~\eqref{PDF-zeta-vs-delta-phi}.  To achieve this, it would be plausible to have a situation where the $\rho_{\delta \phi}$ factor approaches a constant in the same limit. A simple relation like the following fulfills our requirement:
\begin{equation}\label{deltaphi_zeta}
\delta \phi \simeq \frac1\alpha \left[ \zeta_0^{-1} - \left(\zeta + \zeta_c\right)^{-1} \right],
\end{equation}
where $\alpha$, $\zeta_0$ and $\zeta_c$ are constants.  From the last relation it is evident that $\delta\phi$ and hence $\rho_{\delta\phi}$ approaches a constant, as desired.  In fact, we have achieved more: The remaining factor, i.e., $d\delta \phi/d\zeta$ gives a power-law decay of the PDF scaling as $1/\zeta^2$ for large $\zeta$, which is indeed heavy-tailed with ${\cal D} = 2/\zeta \to 0$.  We need to show that there is indeed a potential that supports this solution.  Let us denote the number of $e$-folds required to reach $\phi_2$ from the initial configuration $\phi_1$ by $N(\phi_1, \phi_2)$.\footnote{There is an implicit dependence on the initial velocity $\pi_1 = \phi'_1$ in these relations, which we do not write explicitly since we work with a fixed $\pi_1 = \bar\pi$, ignoring $\delta\pi$ perturbations. See the body of the paper for justification.}
Inverting Eq.~\eqref{deltaphi_zeta} and invoking Eq.~\eqref{deltaN}, this implies
\begin{equation}
-\left( \alpha \delta \phi - \zeta_0^{-1} \right)^{-1} -\zeta_c \simeq N(\bar \phi+\delta \phi, \phi_e)- N(\bar \phi, \phi_e).
\end{equation}
From this, we may identify
\begin{equation}
\zeta_c = N(\bar\phi, \phi_e), \qquad \zeta_0  = \frac1{\alpha \left( \phi_e - \bar \phi \right)}.
\end{equation}
Therefore,
\begin{equation}
N(\bar \phi+\delta \phi, \phi_e) \simeq - \frac1{\alpha \left[ (\delta \phi + \bar\phi) - \phi_e \right]},
\end{equation}
which suggests that
\begin{equation}\label{N_phi}
N(\phi_1, \phi_2) = \frac1{\alpha \left( \phi_2 - \phi_1 \right)}.
\end{equation}
Notice that these expressions were---and still are---approximate (for large $\zeta$) and may not be taken literally.  In particular, Eq.~\eqref{N_phi} is not well defined when $\phi_1 = \phi_2$.  Nonetheless, we can use them to {\it guess} a potential that may work. Assuming that the inflaton field evolves monotonically, we can write Eq.~\eqref{KG} in the following equivalent form
\begin{equation}\label{KG-inf}
\frac{V_{,\phi}}{V} = -\frac{1}{N_{,\phi}} + \frac{N_{,\phi\phi}/N_{,\phi}}{3N_{,\phi}^2 - \frac12}.
\end{equation}
Using Eq.~\eqref{N_phi} for $\phi_1=\bar\phi$ and $\phi_2=\phi(N)$, a simple integration results in
\begin{equation}\label{V1}
V(\phi) = V_0 \left[ 1 + \frac13 \alpha \left( \phi - \bar\phi \right)^3 \right].
\end{equation}
We have made two approximations in deriving this potential.  First, we have ignored the first slow-roll parameter in Eq.~\eqref{KG-inf} ($\frac12 N_{,\phi}^{-2} = \frac12 \phi'^2 = \epsilon \ll 1$ which is valid in an inflationary background) without assuming anything about higher slow-roll parameters.  Secondly, we have assumed $\alpha|\phi-\bar\phi| \ll 1$ and $\alpha|\phi-\bar\phi|^3 \ll 1$ to obtain a simpler form for the potential, although this was unnecessary. 
Notice that the potential depends on $\bar\phi$ (and, in principle, on $\bar\pi$) which are the initial conditions at the initial hypersurface.  This is not problematic; it just means that for a given set of initial conditions a potential of the form \eqref{V1} can lead to a heavy tail.  This concludes our motivation for considering the potential~\eqref{V1}.  We emphasize again that this is based on some guesswork and a dubious reader may choose to accept it only as a suggestion and wait to see if it indeed leads to a heavy tail.

\bibliography{HeavyTail}

\end{document}